\documentclass[11pt]{article}
\usepackage{amsmath,amsthm}
\usepackage{amssymb}
\usepackage{graphicx}

\newcommand{\be}{\begin{equation}}
\newcommand{\ee}{\end{equation}}

\begin{document}

\title{Unstaggered-staggered solitons on one- and two-dimensional
two-component discrete nonlinear Schr\"{o}dinger lattices}

\author{Robert A. Van Gorder$^{1*}$, Andrew L. Krause$^{2}$, Boris A. Malomed$%
^{3}$, and D. J. Kaup$^{4}$\\$^{1}$ Department of Mathematics and Statistics\\
University of Otago\\
P.O. Box 56, Dunedin 9054, New Zealand\\
$^{2}$Mathematical Institute, University of Oxford\\
Andrew Wiles Building, Radcliffe Observatory Quarter\\
Woodstock Road, Oxford OX2 6GG United Kingdom%
\\
$^{3}$Department of Physical Electronics\\
School of Electrical Engineering, Faculty of Engineering \\
and Center for Light-Matter Interaction,\\
Tel Aviv University, P.O.B. 39040, Tel Aviv, Israel\\
$^{4}$Department of Mathematics\\
and Institute for Simulation and Training\\
University of Central Florida, Orlando, FL 32816-1364 USA\\
$^*$Corresponding author. Email: rvangorder@maths.otago.ac.nz }
\maketitle

\begin{abstract}
We study coupled unstaggered-staggered soliton pairs emergent from a system
of two coupled discrete nonlinear Schr\"{o}dinger (DNLS) equations with the
self-attractive on-site self-phase-modulation nonlinearity, coupled by the
repulsive cross-phase-modulation interaction, on 1D and 2D lattice domains.
These mixed modes are of a ``symbiotic" type, as each
component in isolation may only carry ordinary unstaggered solitons. While
most work on DNLS systems addressed symmetric on-site-centered fundamental
solitons, these models give rise to a variety of other excited states, which
may also be stable. The simplest among them are antisymmetric states in the
form of discrete twisted solitons, which have no counterparts in the
continuum limit. In the extension to 2D lattice domains, a natural
counterpart of the twisted states are vortical solitons. We first introduce
a variational approximation (VA) for the solitons, and then correct it
numerically to construct exact stationary solutions, which are then used as
initial conditions for simulations to check if the stationary states persist
under time evolution. Two-component solutions obtained
include (i) 1D fundamental-twisted and twisted-twisted soliton pairs, (ii)
2D fundamental-fundamental soliton pairs, and (iii) 2D vortical-vortical
soliton pairs. We also highlight a variety of other transient dynamical
regimes, such as breathers and amplitude death. The findings apply to modeling binary Bose-Einstein condensates, loaded in a deep lattice potential, with
identical or different atomic masses of the two components, and arrays of
bimodal optical waveguides.\\
\\
keywords: discrete nonlinear Schr\"{o}dinger equations; unstaggered-staggered lattice; variational approximation; solitons\\
\end{abstract}

\section{Introduction}

Discrete nonlinear Schr\"{o}dinger (DNLS) equations provide models for a
great variety of physical systems \cite{Panos}. A well-known implementation
of the basic DNLS equation is provided by arrays of transversely coupled
optical waveguides, as predicted in \cite{Demetri} and realized
experimentally, in various optical settings \cite{Silberberg,
Moti,Jena,Jena2}. A comprehensive review of nonlinear optics in discrete
settings was given by Ref. \cite{review}. Another realization of the DNLS
equation in provided by Bose-Einstein condensates (BECs) loaded into deep
optical-lattice potentials, which split the condensate into a chain of
droplets trapped in local potential wells, which are tunnel-coupled across
the potential barriers between them \cite%
{discreteBECexperiment,discreteBECexperiment2}. In the tight-binding
approximation, this setting is also described by the DNLS version of the
Gross-Pitaevskii (GP) equation \cite{discreteBECtheory,discreteBECtheory2,
discreteBECtheory3, discreteBECtheory4,discreteBECtheory5}.

One-dimensional (1D) DNLS equations with self-attractive and self-repulsive
on-site nonlinearity generate localized modes of \textit{unstaggered} and
\textit{staggered} types, respectively. In the latter case, the on-site
amplitudes alternate between adjacent sites of the lattice \cite{Panos}. In
the continuum limit, the unstaggered discrete solitons carry over into
regular ones, while the staggered solitons correspond to \textit{gap solitons%
}, which are supported by the combination of self-defocusing nonlinearity
and spatially periodic potentials \cite{Brazhnyi,Morsch,gapsol}.

Many physical settings are modeled by systems of coupled DNLS equations. In
optics, they apply to the bimodal propagation of light represented by
orthogonal polarizations or different carrier wavelengths. In BEC, coupled
GP equations describe binary condensates \cite{Pit}. Usually, bimodal
discrete solitons in two-component systems are considered with a single type
of their structure in both components, either unstaggered or staggered,
because the self-phase- and cross-phase-modulation (SPM\ and XPM) terms,
acting in each component and coupled nonlinearly, are assumed to have
identical signs \cite{Panos}. Nevertheless, the opposite signs are also
possible in BEC, where either of them may be switched by means of the
Feshbach resonance \cite{Pit,FR,FR2,FR3,FR4}. Discrete solitons of the
\textit{mixed} type, built as complexes of unstaggered and staggered
components, were introduced in Ref. \cite{we}, assuming opposite SPM and
XPM\ signs. Earlier, single-component states of a mixed
unstaggered-staggered type were investigated in the form of surface modes at
an interface between different lattices \cite{Molina,Molina2}. In continuum
systems, counterparts of mixed modes are represented by \textit{semi-gap}
solitons, which are bound states of an ordinary soliton in one component and
a gap soliton in the other \cite{semi}.

The mixed modes reported in Ref. \cite{we} are \textquotedblleft symbiotic"
ones, as each component in isolation may support solely ordinary unstaggered
solitons. The results were obtained in an analytical form, using the
variational approximations (VA), and verified by means of numerical methods.
It was found that almost all the symbiotic solitons were predicted by the VA
accurately, and were stable. Unstable solitons were found only close to
boundary of their existence region, where the solitary modes have very broad
envelopes, being poorly approximated by the VA.

Most works on DNLS systems concern symmetric on-site-centered fundamental
solitons, which represent the ground state of the corresponding model \cite%
{Panos}, including the unstaggered-staggered solitons \cite{we}.
Furthermore, only fundamental solitons represent stationary states in the
continuum NLS equation. However, DNLS models give rise to stationary excited
states, which may be stable too. The simplest among them are antisymmetric
states in the form of\ discrete \textit{twisted solitons} \cite{twisted},
which have no counterparts in the continuum limit. Once
unstaggered-staggered discrete solitons are possible in two-component DNLS\
systems, it is natural to introduce the twist in the latter setting too,
with three different species of such discrete solitons possible, which are
\textit{single-twisted}, in either component---staggered or unstaggered
one---or \textit{double-twisted}, in both components.

Discrete solitons on two-dimensional (2D) lattices have been studied in a
variety of contexts,  \cite{lederer2008discrete,Panos}. Experimental
literature highlights the existence of real solitons on 2D optically induced
nonlinear photonic lattices \cite{fleischer2003observation,
fleischer2004observation}. A natural extension of the earlier analysis,
performed in the 1D setting \cite{we}, is to build two-component
unstaggered-staggered complexes in 2D two-component DNLS systems, which may
be realized physically in the same physical settings (optics and BEC) as
mentioned above, provided that the corresponding waveguiding arrays are
built, in the transverse plane, as 2D lattices, or the BEC is loaded into a
deep 2D optical-lattice potential. In addition to 2D fundamental discrete
solitons which may be naturally expected in the unstaggered-staggered
system, one may also look for compound modes in which one or both components
are represented by discrete vortex solitons \cite{vortex}, which is the 2D
analogue of the 1D twisted solitons. Vortex solitons have previously been
found in various optical setups \cite{yang2003fundamental,
musslimani2004self, ferrando2004vortex}, see also recent reviews \cite%
{Astra,PhysD}. Experimental creation of discrete vortex solitons in
self-focusing optically induced lattices was reported in Refs. \cite%
{fleischer2004observation, neshev2004observation}.

The remainder of this paper is organized according to the dimension of the
lattice domain, with single- and double-twisted two-component
unstaggered-staggered solitons on 1D lattices considered in Section \ref%
{sec2}. The analysis of unstaggered-staggered 2D discrete soliton complexes,
consisting of fundamental soliton pairs, along with the more sophisticated
fundamental-vortical and vortical-vortical ones, is reported in Section \ref%
{sec3}. For both dimensions of the lattice, we present the governing DNLS
equations and the corresponding Lagrangian. Assuming a decay rate for the
solitons' tails as predicted by the linearization of the DNLS equations, we
elaborate the variational approximation (VA) for each type of soliton. We
then use the VA-produced predictions as an initial guess to obtain the
corresponding states in a numerically exact form. This approach is useful,
as without an appropriate input the numerical scheme may readily converge to
zero or some non-physical state. Furthermore, for stationary states which
are stable and symmetric between the components (fundamental-fundamental,
twist-twist, or vortical-vortical), the agreement of VA with numerical
findings is quite good, whereas in the case of asymmetric pairs of the
components (one twisted or vortical, the other being fundamental) the
agreement is less accurate. Starting with numerically exact stationary states,
we then simulate their evolution in time to determine what states are stable
or unstable. Concluding remarks are made in Section \ref{sec4}.

\section{One-dimensional coupled unstaggered-staggered modes}

\label{sec2} In this section, we initiate the analysis by formulating VA,
which has proved to be quite efficient in the studies of fundamental
discrete solitons in diverse settings, as shown at heuristic \cite%
{MW1996,MW19962,VA,VA2,VA3,VA4,VA5, VA6,VA7,VA8,VA9,VA10,VA11,VA12,
VA13,VA14,VA15,VA16,VA17,VA18,VA19,VA20,VA21, VA22,VA23,VA24,VA25,VA26} and
more rigorous \cite{Schneider,Schneider2} levels. We introduced the DNLS
equations and their Lagrangian in subsection \ref{1dframework}, elaborate
the VA in subsection \ref{1dva}, and report results for numerically exact
stationary solutions and their subsequent temporal evolution in subsection %
\ref{1dna}. We further select stable stationary states in subsection \ref%
{1dstab}, and then outline transient regimes related to the evolution of
unstable modes in subsection \ref{1dtrans}.

\subsection{The DNLS equations}

\label{1dframework} In Ref. \cite{we}, a system of coupled DNLS equations
for discrete fields $\phi _{n}$ and $\psi _{n}$ was introduced in 1D:
\begin{subequations}
\label{phipsi}
\begin{eqnarray}
i\frac{d}{dt}\phi _{n} &=&-\frac{1}{2}\left( \phi _{n+1}+\phi _{n-1}-2\phi
_{n}\right) -\left( \left\vert \phi _{n}\right\vert ^{2}+\beta \left\vert
\psi _{n}\right\vert ^{2}\right) \phi _{n}, \\
i\frac{d}{dt}\psi _{n} &=&-\frac{1}{2M}\left( \psi _{n+1}+\psi _{n-1}-2\psi
_{n}\right) -\left( \left\vert \psi _{n}\right\vert ^{2}+\beta \left\vert
\phi _{n}\right\vert ^{2}\right) \psi _{n},
\end{eqnarray}%
where $M>0$ is the relative atomic mass of the two species in the case of
BEC, or the inverse ratio of the intersite coupling constants in the
waveguide array, and $\beta <0$ is the relative coefficient of the onsite
XPM coupling between the fields, assuming that coefficients of the
self-attractive SPM nonlinearity for both fields are scaled to be $1$.

Solutions with unstaggered $\phi _{n}$ and staggered $\psi _{n}$
onsite-centered components and two chemical potentials, $\lambda $ and $\mu $%
, are sought for as
\end{subequations}
\begin{equation}
\phi _{n}(t)=e^{-i\lambda t}u_{n},~\psi _{n}(t)=e^{-i\mu t}\left( -1\right)
^{n}v_{n},  \label{stag}
\end{equation}%
where real discrete fields $u_{n}$ and $v_{n}$ satisfy the following
stationary equations,
\begin{subequations}
\label{uv}
\begin{eqnarray}
\left( \lambda -1\right) u_{n}+\frac{1}{2}\left( u_{n+1}+u_{n-1}\right)
+\left( u_{n}^{2}+\beta v_{n}^{2}\right) u_{n} &=&0, \\
\left( \mu -\frac{1}{M}\right) v_{n}-\frac{1}{2M}\left(
v_{n+1}+v_{n-1}\right) +\left( v_{n}^{2}+\beta u_{n}^{2}\right) v_{n} &=&0,
\end{eqnarray}%
which can be derived from the\ Lagrangian,
\end{subequations}
\begin{equation}
L=\frac{1}{2}\sum_{n=-\infty }^{+\infty }\left[ -\frac{1}{2}\left(
u_{n+1}-u_{n}\right) ^{2}+\lambda u_{n}^{2}+\frac{1}{2M}\left(
v_{n+1}-v_{n}\right) ^{2}+\left( \mu -\frac{2}{M}\right) v_{n}^{2}+\frac{1}{2%
}u_{n}^{4}+\frac{1}{2}v_{n}^{4}+\beta u_{n}^{2}v_{n}^{2}\right] .
\label{Lagr}
\end{equation}
In the standard way, the respective equations \eqref{uv} can be obtained by taking the first variation of \eqref{Lagr} in $u_n$ and $v_n$, respectively.

Note that, in the limit of $M\gg 1$, which is tantamount to the Thomas-Fermi
(TF)\ approximation for discrete equation (\ref{uv}b), this equation allows
one to eliminate $v_{n}$ in favor of $u_{n}$,%
\begin{equation}
v_{n}^{2}=-\mu -\beta u_{n}^{2}\,,  \label{v^2}
\end{equation}%
hence in this case the coupled stationary system (\ref{uv})\ reduces to a
single equation. In the opposite limit of $M\ll 1$, large constant $M^{-1}$
makes Eq. (\ref{uv}b) close to its continuum counterpart, where $n$ may be
considered as a continuous coordinate, and, accordingly, the broad $v(n)$
component interacts with a narrow strongly discrete one, $u_{n}$. Thus, the
effective continuous equation for $v(n)$ amounts to%
\begin{equation}
\mu v-\frac{1}{2M}\frac{d^{2}v}{dn^{2}}+v^{3}+\beta W_{u}\delta \left(
n\right) v=0,  \label{cont}
\end{equation}%
where $\delta (n)$ is the delta-function, and $W_{u}$ is the norm of the $u$%
-component, $W_{u}=\sum_{n=-\infty }^{\infty }u_{n}^{2}$. In the same
approximation, $v_{n}^{2}$ in Eq. (\ref{uv}a) reduces to a constant, $%
v_{n}\approx v\left( n=0\right) $. For $\beta <0$, which is considered in
this work, Eq. (\ref{cont}) readily gives rise to an exact solution in the
form of a pinned soliton,%
\begin{equation}
v(n)=\frac{\sqrt{2\mu }}{\sinh \left( \sqrt{2\mu M}\left( |n|+n_{0}\right)
\right) },  \label{sing}
\end{equation}%
with $n_{0}$ determined by condition%
\begin{equation}
\coth \left( \sqrt{2\mu M}n_{0}\right) =-\beta W_{u}\sqrt{M/(2\mu )}.
\label{coth}
\end{equation}%
As follows from Eq. (\ref{coth}), this continuum-limit solution exists in
the following interval of chemical potential of the staggered component: $%
0<\mu <\left( M/2\right) \left( \beta W_{u}\right) ^{2}$.

\subsection{The variational approximation}

\label{1dva} In the general case, asymptotic tails of stationary discrete
solitons decay at $|n|\rightarrow \infty $ as
\begin{equation}
u_{n}=Ae^{-p|n|},\quad v_{n}=Be^{-q|n|},  \label{ansatz}
\end{equation}%
with $p$ and $q$ determined by the linearized limit of Eqs. (\ref{uv}) at $%
n\rightarrow \pm \infty $. Substituting expressions \eqref{ansatz} in the
linearized equations, we find $\lambda -1+\cosh (p)=0$ and $\mu -\frac{1}{M}-%
\frac{1}{M}\cosh (q)=0$. Solving these equations for $p$ and $q$ in terms of
$\lambda $, $\mu $, and $M$, we obtain
\begin{subequations}
\label{pq}
\begin{eqnarray}
p &=&\text{arccosh}\left( 1-\lambda \right) \equiv \ln \left( 1-\lambda +%
\sqrt{-\lambda \left( 2-\lambda \right) }\right) ,~ \\
q &=&\text{arccosh}\left( M\mu -1\right) \equiv \ln \left( M\mu -1+\sqrt{%
M\mu \left( M\mu -2\right) }\right) \,.
\end{eqnarray}%
For $p$ and $q$, given by Eqs. (\ref{pq}) to be real and positive, the
allowed ranges of chemical potentials $\mu $ and $\lambda $ are
\end{subequations}
\begin{equation}
\lambda <0\,,\quad \mu >\frac{2}{M}\,.  \label{><}
\end{equation}

Fundamental-soliton solutions of Eqs. \eqref{phipsi} were constructed in
Ref. \cite{we}, while in the present section we extend those results to more
sophisticated solutions. A new possibility is to look for \textit{twisted}
solitons (i.e., spatially-antisymmetric ones). In the single-component DNLS
equation, twisted solitons were introduced in Ref. \cite{twisted}. An
initial ansatz for such solitons, with the twisted structure in one or both
components, can be taken as
\begin{eqnarray}
u_{n} &=&Ae^{-p|n|},\quad v_{n}=Bne^{-q|n|},  \label{01} \\
u_{n} &=&Ane^{-p|n|},\quad v_{n}=Be^{-q|n|},  \label{10} \\
u_{n} &=&Ane^{-p|n|},\quad v_{n}=Bne^{-q|n|},  \label{11}
\end{eqnarray}%
where $p$ and $q$ are produced by the solution of the linearized equations
for the tails in the same form of Eq. (\ref{pq}) as above. Note that, if the
$u$-component is twisted, while the $v$ mode is not, then the TF\
approximation based on Eq. (\ref{v^2}) is definitely irrelevant, as it would
predict $v_{0}^{2}=-\mu <0$ at $n=0$. We stress that the two composite
straight-twisted configurations, corresponding to \textit{ans\"{a}tze} (\ref%
{01}) and (\ref{10}), are not equivalent, as in these cases the twisted
constituent is created in the unstaggered or staggered component,
respectively.

Studies of stability of plane waves in 1D DNLS lattices suggest that these
states are often modulationally unstable to small-wavenumber perturbations
\cite{kivshar1992modulational,kobyakov1998dark, jin2005modulational}.
Perturbations of unstable plane waves of this type often result in the
emergence of highly localized structures, such as solitons or other solitary
waves \cite{kivshar1992modulational, kivshar1993creation}. In this case,
solitons of the form given by Eqs. \eqref{01} and \eqref{11} are expected to
have narrow envelopes, meaning large values of $p$ and $q$. More broad
envelopes are likely to be modulationally unstable against planewave 
perturbations in the large wavenumber regime. 
This intuition is in agreement with the results of Ref. \cite%
{we} for fundamental solitons, and we will later show that the stability of
solutions that we find here agrees with those earlier findings, i.e., stable
solutions exhibit rapidly decaying tails, while broader soliton envelopes
tend to be unstable.

Assuming the presence of twisted unstaggered and straight staggered
components, Eq. \eqref{10} gives $u_{0}=0$, $u_{-n}=-u_{n}$, $v_{-n}=v_{n}$,
and $v_{0}>v_{1}>v_{2}>\dots $. For $n=0$, the $u$-component of Eq. %
\eqref{uv} is identically satisfied, whereas the equation for $v$ yields $%
\frac{v_{1}}{v_{0}}=M\mu -1+Mv_{0}^{2}$. As the ratio $v_{1}/v_{0}$ must
satisfy restriction $0<v_{1}/v_{0}<1$ in localized solutions, this implies $%
\frac{1}{M}-\mu <v_{0}^{2}<\frac{2}{M}-\mu $. However, condition $\mu >2/M$
from Eq. (\ref{><}) then results in $v_{0}^{2}<0$, therefore there may be no
real amplitude $v_{0}$ generating twisted unstaggered and straight staggered
components. This does not mean that other solutions of such a type (perhaps
non-stationary ones) do not exist, but solely that solitons do not exist in
this form. Thus, we exclude solitons of the form given by Eq. \eqref{10}
from the consideration, and we will only seek solitons of the form defined
by Eqs. \eqref{01} and \eqref{11}. For each of these, we will obtain
analytical approximations through the VA, which will be verified by means of
numerical methods.

Now, we outline the construction of solitons of the form \eqref{01} and %
\eqref{11} by way of the VA. The existence of VA solutions is tied to the
existence of a positive solution for squared amplitudes $\left(
A^{2},B^{2}\right) $ of the respective \textit{ans\"{a}tze}, as produced by
the Euler-Lagrange equations, derived in the framework of VA. In turn, we
find that the existence of such positive solutions strongly depends on
values the XPM coefficient, $\beta <0$, and relative mass of the two
components, $M>0$. For a fixed set of these values, it is possible to
determine a subset of the plane $(\lambda ,\mu )$ of the chemical potentials
of the two components, where positive solutions for $\left(
A^{2},B^{2}\right) $ can be found.

\subsubsection{Fundamental unstaggered - twisted staggered pairs}

We start the VA analysis for \textit{ans\"{a}tze} (\ref{01}). Substituting it in
Lagrangian (\ref{Lagr}) and performing the summation yields the following
effective Lagrangian:
\begin{eqnarray}
2L_{\mathrm{eff}} &=&-A^{2}\tanh (p/2)+\lambda A^{2}\coth p+\frac{A^{4}}{2}%
\coth (2p)+\frac{B^{2}}{2M}\frac{\left( \sinh q\right) ^{-1}}{\cosh (q)+1}
\notag \\
&&+\frac{B^{2}}{2}\left( \mu -\frac{2}{M}\right) \left( \coth ^{3}q-\coth
q\right) +\frac{B^{4}}{2}\left[ \frac{3}{2}\coth ^{5}(2q)-\frac{5}{2}\coth
^{3}(2q)+\coth (2q)\right]   \notag \\
&&+\frac{\beta A^{2}B^{2}}{2}\left[ \coth ^{3}(p+q)-\coth (p+q)\right] \,,
\label{Leff01}
\end{eqnarray}%
which gives rise to the Euler-Lagrange equations,
\begin{equation}
\frac{\partial L_{\mathrm{eff}}}{\partial \left( A^{2}\right)} = \frac{\partial L_{\mathrm{eff}}}{\partial \left( B^{2}\right)} =0.  \label{EL}
\end{equation}%
They amount to a system of linear equations for $A^{2}$ and $B^{2}$:
\begin{equation}
A^{2}\left[ \coth (2p)\right] +\frac{\beta }{2}B^{2}\left[ \coth
^{3}(p+q)-\coth (p+q)\right] =\tanh (p/2)-\lambda \coth p\,,  \label{var01a}
\end{equation}%
\begin{eqnarray}
&&\frac{\beta }{2}A^{2}\left[ \coth ^{3}(p+q)-\coth (p+q)\right] +B^{2}\left[
\frac{3}{2}\coth ^{5}(2q)-\frac{5}{2}\coth ^{3}(2q)+\coth (2q)\right]
\notag \\
&&\qquad =-\frac{1}{2M}\frac{\left( \sinh q\right) ^{-1}}{\cosh q+1}-\frac{1%
}{2}\left( \mu -\frac{2}{M}\right) \left( \coth ^{3}q-\coth q\right) \,.
\label{var01b}
\end{eqnarray}
Note that $p$ and $q$ are already determined from \eqref{pq}. The system \eqref{var01a}-\eqref{var01b} is solved for the unknown $A$ and $B$, which are the initial amplitudes of the variational approximation at $n=0$. As the system is linear in $A^2$ and $B^2$, it is sufficient to obtain a solution for the quantities $A^2$ and $B^2$, and these quantities must be positive. We then take the positive root $\sqrt{A^2}=A$, $\sqrt{B^2}=B$ for the initial amplitudes.

Physically relevant solutions to Eqs. (\ref{var01a}) and (\ref{var01b}),
with $A^{2}>0$ and $B^{2}>0$, do not exist for $\beta >0$, but they may
exist if $\beta <0$, i.e., in the case of the repulsive interaction between
the two components, which is the subject of the present work. The linear
system of Eqs. \eqref{var01a}-\eqref{var01b} becomes degenerate (with zero
determinant) in the case of
\begin{equation}
\coth (2p)\left[ \frac{3}{2}\coth ^{5}(2q)-\frac{5}{2}\coth ^{3}(2q)+\coth
(2q)\right] =\frac{\beta ^{2}}{4}\left[ \coth ^{3}(p+q)-\coth (p+q)\right]
^{2}\,.  \label{degen01}
\end{equation}

\subsubsection{Twisted unstaggered - twisted staggered pairs}

Substituting ansatz (\ref{11}) into Lagrangian (\ref{Lagr}) and carrying out
the summation as above, we obtain the effective Lagrangian in the following
form:
\begin{eqnarray}
2L_{\mathrm{eff}} &=&-\frac{A^{2}}{2}\frac{\left( \sinh p\right) ^{-1}}{%
\cosh p+1}+\lambda \frac{A^{2}}{2}\left( \coth ^{3}p-\coth p\right) +\frac{%
B^{2}}{2m}\frac{\left( \sinh q\right) ^{-1}}{\cosh (q)+1}+\frac{B^{2}}{2}%
\left( \mu -\frac{2}{M}\right) \left( \coth ^{3}q-\coth q\right)   \notag \\
&&+\frac{A^{4}}{2}\left[ \frac{3}{2}\coth ^{5}(2p)-\frac{5}{2}\coth
^{3}(2p)+\coth (2p)\right] +\frac{B^{4}}{2}\left[ \frac{3}{2}\coth ^{5}(2q)-%
\frac{5}{2}\coth ^{3}(2q)+\coth (2q)\right]   \notag \\
&&+\beta A^{2}B^{2}\left[ \frac{3}{2}\coth ^{5}(p+q)-\frac{5}{2}\coth
^{3}(p+q)+\coth (p+q)\right] \,.  \label{Leff11}
\end{eqnarray}%
It is convenient to define
\begin{equation}
\chi (\alpha )\equiv \frac{3}{2}\coth ^{5}\alpha -\frac{5}{2}\coth
^{3}a+\coth \alpha ,  \label{chi}
\end{equation}%
which is positive for all $\alpha >0$. Euler-Lagrange equations (\ref{EL})
following from Lagrangian (\ref{Leff11}) can be written as
\begin{equation}
A^{2}\chi (2p)+\beta B^{2}\chi (p+q)=\frac{1}{2}\frac{\left( \sinh p\right)
^{-1}}{\cosh (p)+1}-\frac{\lambda }{2}\left( \coth ^{3}(p)-\coth (p)\right)
\,,  \label{var11a}
\end{equation}%
\begin{equation}
\beta A^{2}\chi (p+q)A^{2}+B^{2}\chi (2q)=-\frac{1}{2M}\frac{\left( \sinh
q\right) ^{-1}}{\cosh (q)+1}-\frac{1}{2}\left( \mu -\frac{2}{M}\right)
\left( \coth ^{3}(q)-\coth (q)\right) \,.  \label{var11b}
\end{equation}%
The system \eqref{var11a}-\eqref{var11b} is again a system for unknown initial amplitudes $A$ and $B$ in the variational approximation. Once again, this system may give rise to physical solutions only in the case
of the repulsive XPM interaction, $\beta <0$. The system of Eqs. %
\eqref{var11a} and \eqref{var11b} becomes degenerate if
\begin{equation}
\chi (2p)\chi (2q)=\beta ^{2}\chi ^{2}(p+q)\,,  \label{degen11}
\end{equation}%
where $\chi $ is defined as per Eq. (\ref{chi}). Note that $\chi (\alpha
)\rightarrow \infty $ as $\alpha \rightarrow +0$ and $\chi (\alpha
)\rightarrow 0$ as $\alpha \rightarrow \infty $, and that $\chi $ is a
decreasing function in its domain. Therefore, the ratio $\chi (2p)\chi (2q)/\chi
(p+q)$ is always positive, and there always exists a value
\begin{equation}
\beta =\overline{\beta }\equiv -\sqrt{\frac{\chi (2p)\chi (2q)}{\chi (p+q)}}%
<0\,,
\end{equation}%
at which system \eqref{var11a}-\eqref{var11b} is degenerate.

\subsection{The numerical approach}

\label{1dna} Localized steady-state solutions were numerically computed by
solving a truncated form of Eq. \eqref{uv} on a finite lattice. This was implemented with periodic boundary conditions, although we also compared results for fixed homogeneous conditions to ensure the boundaries played no role in the localized solution structure. The nonlinear system was then solved via the Matlab function `\emph{fsolve},' which implements a Newton-like solution procedure (specifically a trust-region dogleg method). Unless otherwise mentioned, these solutions were carried out using $10^{3}+1$ lattice points, though they are nearly identical to solutions on much smaller lattices,
hence it is safe to assume that the truncation has not changed the structure
of localized solutions. We fixed function and optimality tolerances of $%
10^{-12}$, and always checked that both the vector ($L^{2}$) and component ($%
L^{\infty }$) norms of the objective function, evaluated on the numerical
solution, were small enough ($<10^{-4}$, and typically much smaller). The VA
solitons were used as the initial guess. In some cases the resultant
numerical solution was close to the original VA prediction, while in other
cases, the initial VA solution converged to a soliton where one component is
zero. In the latter case, to find nontrivial solutions in both components,
we deflated the objective function away from zero (see \cite%
{charalampidis2018computing} for a similar technique applied to the 2D GP
equation). Specifically, if $\mathbf{F}$ is the objective function (the
left-hand side of Eq. \eqref{uv}), we replace it by
\begin{equation}
\left( 1+\frac{1}{\max (|u_{n}|)}\right) \left( 1+\frac{1}{\max (|v_{n}|)}%
\right) \mathbf{F}.
\end{equation}

Once an exact stationary state is found, we tested its stability by
simulations of their evolution in the framework of Eq.~\eqref{phipsi}. It
was thus found that some stationary solutions are unstable, as they do not
persist in the course of the evolution, instead breaking apart or evolving
into solutions of other kinds. Many solutions which do stably persist, emit
a small amount of radiation at the initial stage of the evolution, whilst
the solution adjusts to the true steady state. Unstable states tend to
gradually break up into radiation, which disperses throughout the domain. We
define a solution as stable if both $\phi _{n}$ and $\psi _{n}$ components
persist, keeping constant absolute values

To run simulations of the evolution, we set $\phi _{n}(0)=u_{n}$ and $\psi
_{n}(0)=v_{n}$, and evolved Eqs. \eqref{phipsi} using the Matlab function `%
\emph{ode45}', which implements a fourth-order Runge-Kutta scheme, as a
variation of an algorithm from Ref. \cite{dormand1980family}. We fixed
absolute and relative tolerances of $10^{-13}$. The solutions were
consistent with simulations computed with the help of the first-order stiff solver, `%
\emph{ode15s},' for the simulation time scales, therefore we do not
anticipate accumulations of errors due to round-off or loss of mass for the
duration of the simulations. The simulations terminated when reflection from
the boundary occurred, to avoid artifacts caused by the reflected waves.
Thus, increasing the lattice size, we may increase the simulation time, and
from this we conclude that the solutions we claim to be stable will be
stable on an infinite lattice.

\subsection{Stable 1D two-component soliton}

\label{1dstab} For $\beta =-0.5>-1$, changes in the
relative-mass parameter $M$ primarily shift regions in the $(\mu ,\lambda )$
plane which admit such solutions, but roughly preserve their geometry. In
contrast, increasing $M$ for $\beta =-5<-1$ leads to an increase of the
region of admissible localized solutions. For $\beta =-0.5$, the existence
regions are similar to those found for fundamental soliton pairs on
unstaggered-staggered lattices in Ref. \cite{we} (not shown here in detail).

In fact, most parameter values and initial guesses led to unstable time
evolution, but large stability regions for two-component solitons of the
twisted-twisted soliton were found at large values of $-\lambda $ and $\mu $
(for $\beta =-0.5>-1$). We have also found stable solitons close to but
somewhat different from inputs in the form of Eqs. \eqref{01} or \eqref{11},
such as states with broad multiple-site peaks. In contrast, two-component
solitons of the fundamental-twisted type are, generally, less stable, due to
fragility of such solutions to time dynamics. Thus, we expect that it may be
more difficult to create fundamental-twisted soliton pairs in the
experiment, whereas the robust twisted-twisted pairs should be available for
a variety of parameter regimes and experimental configurations.

\begin{figure}
\centering
\includegraphics[width=0.32\textwidth]{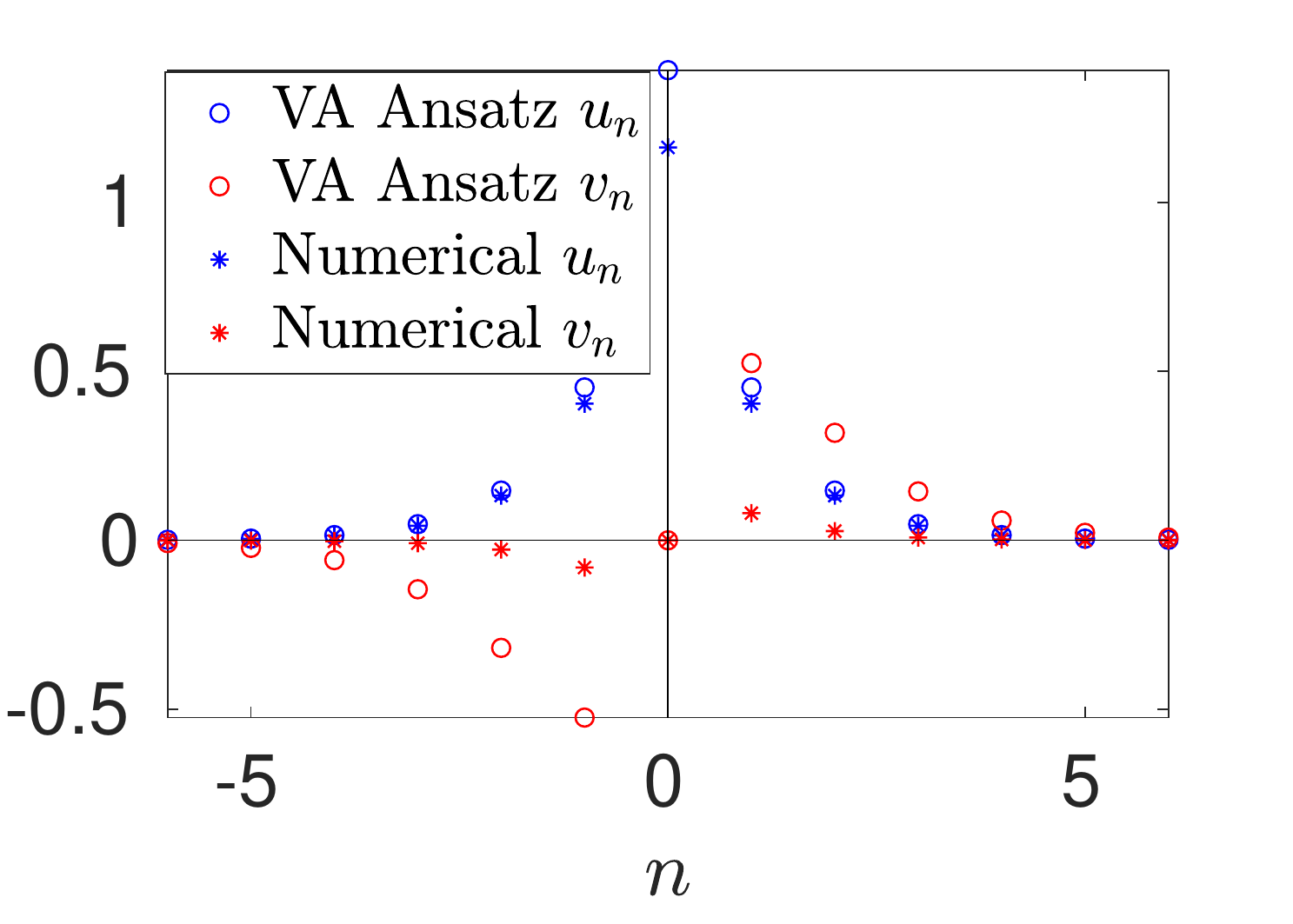} \includegraphics[width=0.32%
\textwidth]{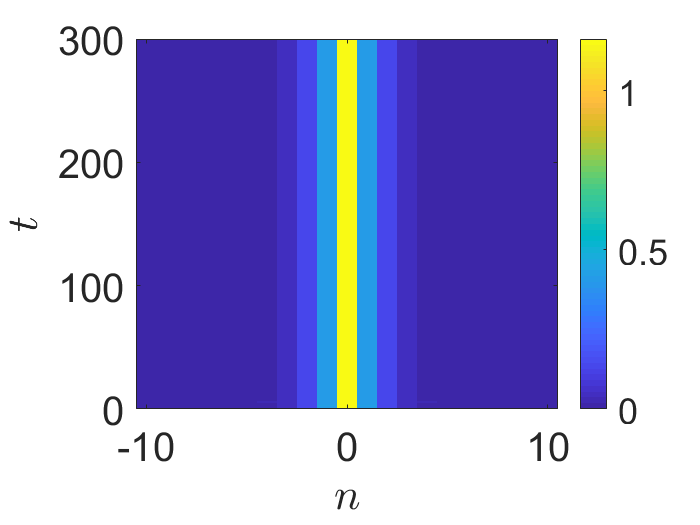} \includegraphics[width=0.32\textwidth]{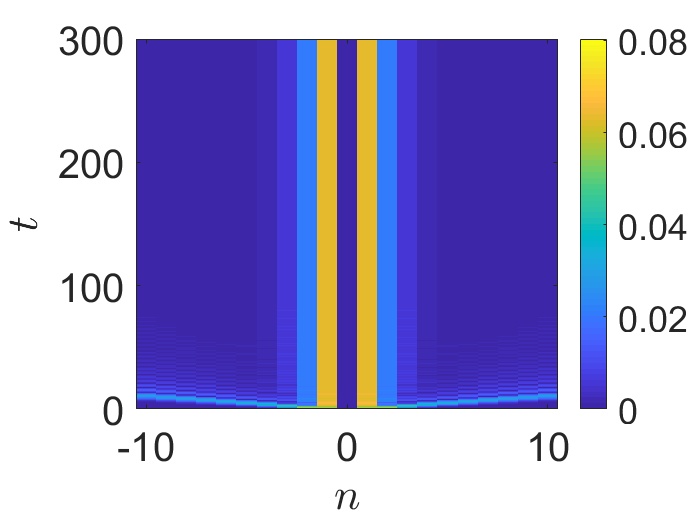}
\par
(a) \hspace{5cm} (b) \hspace{5cm} (c)
\caption{Plots of steady states and time evolution for a stable soliton
built of bound fundamental and twisted components. Parameters are $\protect%
\beta =-10$, $\protect\mu =2.8$, $\protect\lambda =-0.7$, $M=1$. In (a) both
the VA-predicted and numerically exact stationary states are plotted. The
evolution of the two components is displayed in (b) and (c) for $|\protect%
\phi _{n}(t)|$ and $|\protect\psi _{n}(t)|$, respectively. Component $%
\protect\psi _{n}$ shown in (c) initially releases a small amount of
radiation, as it adjusts from the initial configuration to the stable
stationary state.}
\label{stab1a}
\end{figure}

\begin{figure}
\centering
\par
\includegraphics[width=0.32\textwidth]{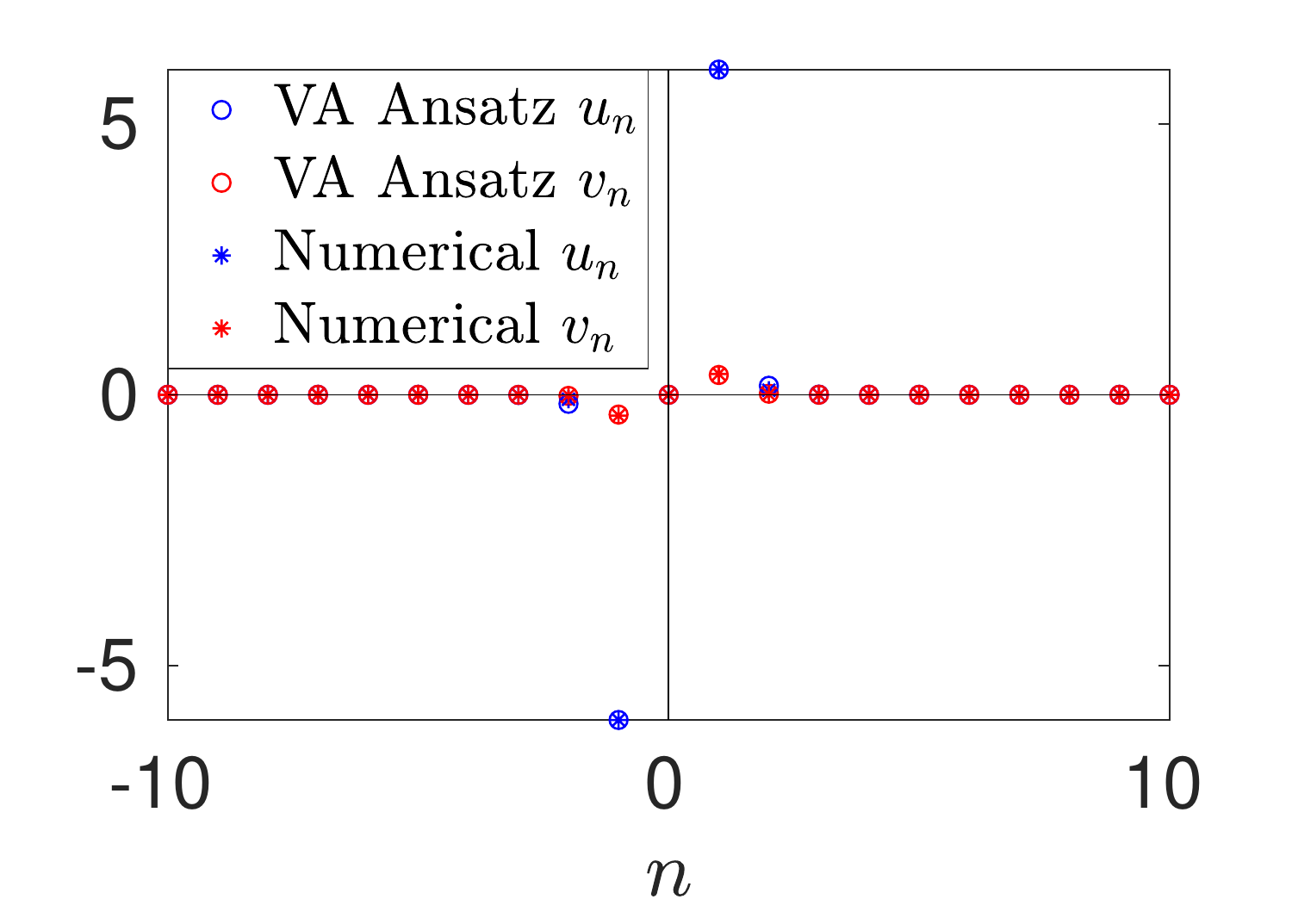} \includegraphics[width=0.32%
\textwidth]{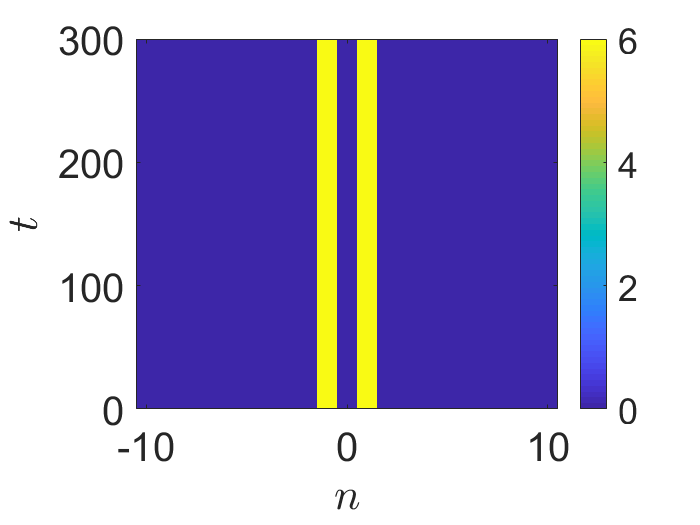} \includegraphics[width=0.32\textwidth]{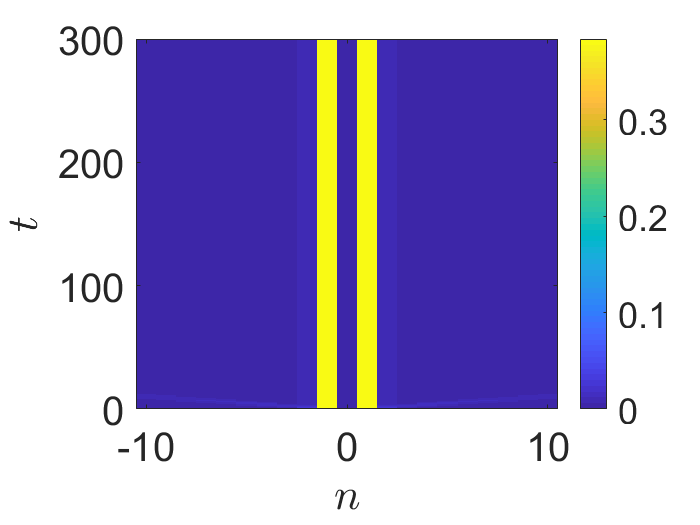}
\par
(a) \hspace{5cm} (b) \hspace{5cm} (c)
\caption{The same as in Fig. \protect\ref{stab1a}, but for a stable solitons
with components of the twisted-twisted type. The parameters are $\protect%
\beta =-0.5$, $\protect\mu =18.9$, $\protect\lambda =-35$, $M=1$. In panel
(c), component $\protect\psi _{n}$ initially releases a small amount of
radiation, as it adjusts from the initial state to the proper stable
stationary one. }
\label{stab1b}
\end{figure}

In other cases, we have found that, in the direct time simulations, unstable
VA solitons with one or two twisted components spontaneously transform into
states with two or one twisted components, respectively. This result
highlights the fact that the VA approach is most useful for detecting highly
localized waves, yet accurate identification of the structure of such states
should be done numerically.

We plot representative stable fundamental-twisted and twisted-twisted
soliton pairs in Figs. \ref{stab1a} and \ref{stab1b}, respectively. The
fundamental-twisted pair shows quantitative disagreement between the VA and
exact numerical solution. Again, as the fundamental-twisted pair is not
robust in the parameter space, the VA based in the simplest ansatz is not a
particularly good fit. In particular, the amplitude of the twisted $v_{n}$
component, shows quantitative disagreement with the VA prediction, whereas
the fundamental components agree with VA quite well. Still, the peaks for
both the VA-predicted twisted component and its numerically found
counterpart are located at the same sites, and it is the amplitude which is
poorly approximated. Simulations of the numerically exact
fundamental-twisted pair demonstrate its stability, after shedding transient
radiation at the initial stage of the evolution.

The twisted-twisted soliton pair shown in Figure \ref{stab1b} features good
agreement between the VA and numerical solutions, which are
indistinguishable at some sites. This fact, along with results for the
fundamental-fundamental soliton pairs displayed in Ref. \cite{we} suggests
that the VA produces accurate predictions when both components have the same
symmetry.

\begin{figure}
\centering
\par
\includegraphics[width=0.4\textwidth]{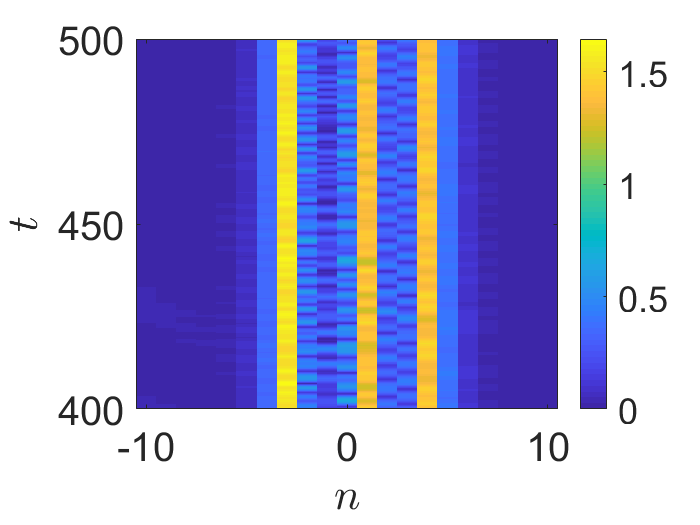} \includegraphics[width=0.4%
\textwidth]{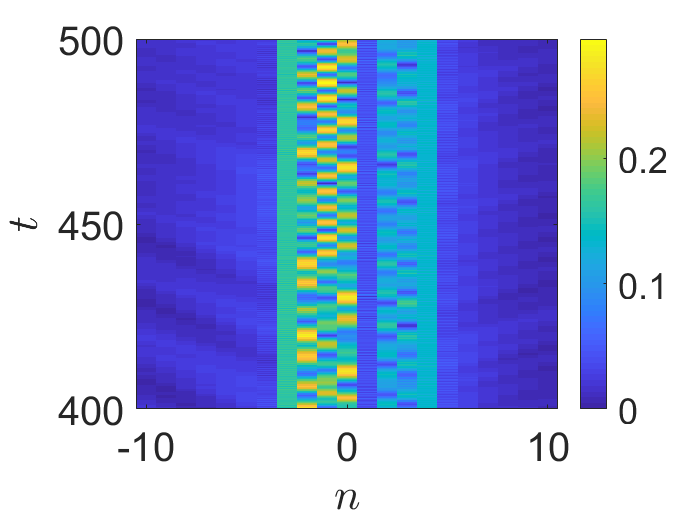}
\par
(a) \hspace{6cm} (b)
\par
\includegraphics[width=0.4\textwidth]{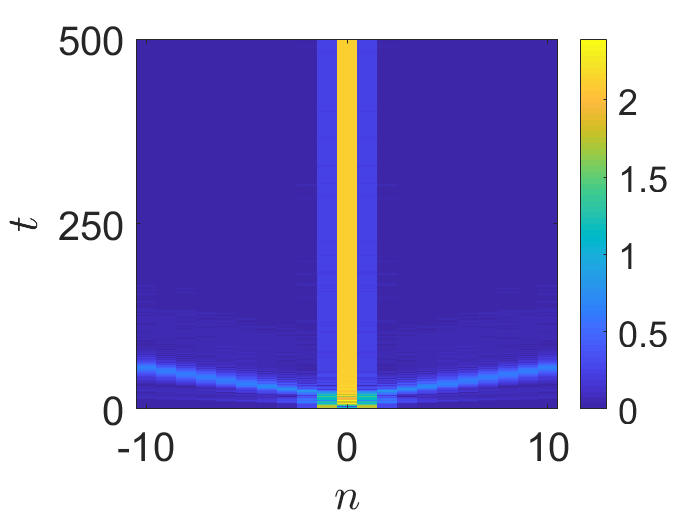} \includegraphics[width=0.4%
\textwidth]{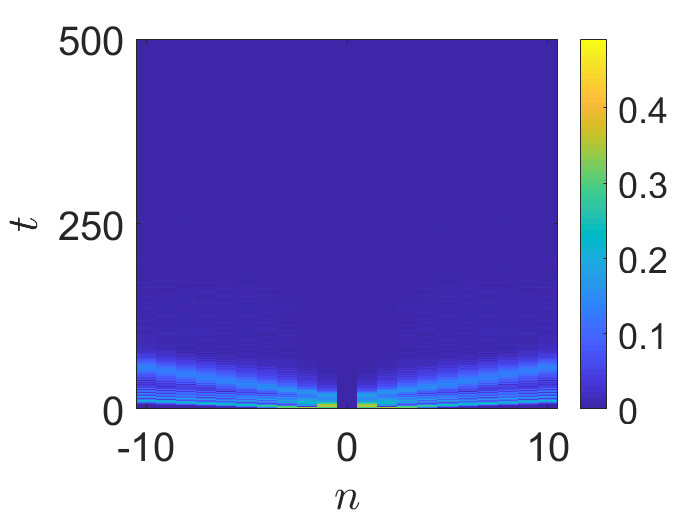}
\par
(c) \hspace{6cm} (d)
\par
\includegraphics[width=0.4\textwidth]{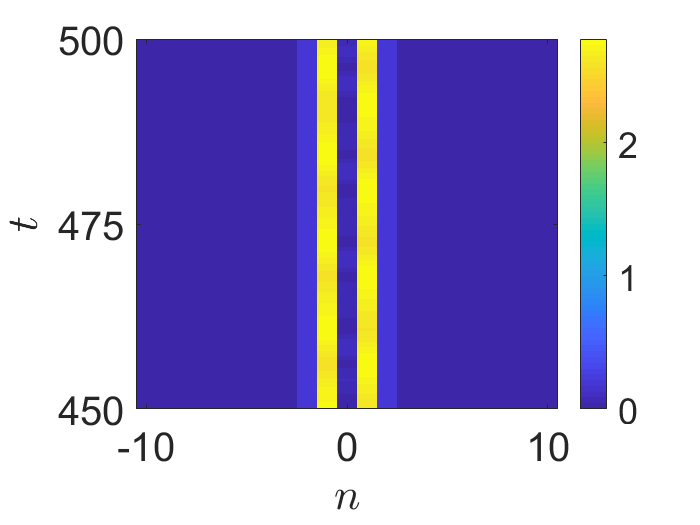} \includegraphics[width=0.4%
\textwidth]{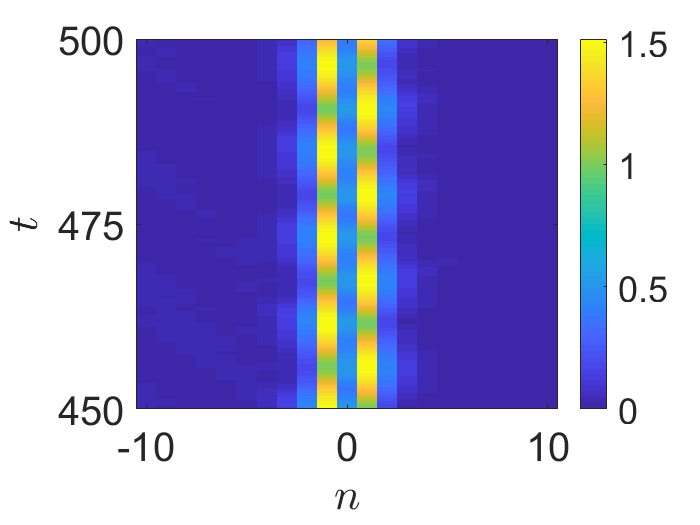}
\par
(e) \hspace{6cm} (f)
\caption{The simulated evolution of unstable numerical found stationary
states, showing different routes of the instability development and eventual
wave breakup. We plot $|\protect\phi _{n}(t)|$ in (a,c,e), and $|\protect%
\psi _{n}(t)|$ in (b,d,f), for parameters (a,b) $\protect\beta =-5$, $%
\protect\mu =5$, $\protect\lambda =-0.5$, $M=1$ in (a,b), $\protect\beta %
=-0.5$, $\protect\mu =2.3$, $\protect\lambda =-2$, $M=1$ in (cd), and $%
\protect\beta =-0.5$, $\protect\mu =2.3$, $\protect\lambda =-5$, $M=1$ in
(e,f). Panels (a,b) exhibit apparent spatiotemporal chaos, confined to a
narrow band. In (c,d) we show a case where one of the components becomes
unstable, decaying into radiation, whereas the other component persists.
After transient emission of radiation, the solution in (e,f) displays
spatiotemporal dynamics akin to a breather. }
\label{inst1}
\end{figure}

\subsection{Transient dynamics}

\label{1dtrans} We have found a variety of evolution routes for unstable
stationary states, as shown in Fig. \ref{inst1}. First, in Fig. \ref{inst1}%
(a) we observe unstable solutions which decay into radiation at large times,
with the core region gradually spreading out over the entire spatial domain
(though some fluctuations persist for large times). In contrast to the
instability resulting in the decay of the original stationary state, there
are other unstable solutions which persist with a finite amplitude in a
finite core region, exhibiting spatiotemporal chaos within it (see Fig.~\ref{inst1}(b)). Spatiotemporal chaos has been observed in other lattice NLS
systems, often under the action of temporal forcing \cite{hennig1999periodic}
or nonlocality \cite{korabel2007transition}. Chaos has also been observed in
DNLS systems which involve nonlinear coupling between adjacent sites \cite%
{ablowitz2000integrability}. Our model does not include any of these
ingredients, with the only change from the standard DNLS equation being the
unstaggered-staggered structure, which gives rise to the increased
complexity in comparison with the standard DNLS lattices.

Further, we have also found solutions featuring unstable dynamics in one
component and apparently stable evolution in the other, as shown in Figs. %
\ref{inst1}(c,d). In particular, this demonstrates \textit{amplitude death}
(vanishing) of one component and persistence of the other. The amplitude
death in lattice dynamical systems modelling many coupled oscillators \cite%
{mirollo1990amplitude, atay2003distributed, konishi2003amplitude}, whereas
the amplitude death in coupled continuum complex Ginzburg-Landau systems has
been demonstrated in Ref. \cite{van2018coupled} for saturable kinetics and
in Ref. \cite{van2019amplitude} for more general yet monotone kinetics. On
the other hand, it was shown in Ref. \cite{van2019amplitude} that coupled
continuum NLS systems do not admit amplitude death. To the best of our
knowledge, the amplitude death regimes has not been found in standard DNLS
systems, again highlighting the rich variety of dynamics possible in
unstaggered-staggered lattices.

In Figs. \ref{inst1}(e,f) we plot solution pairs which maintain their
overall envelope yet exhibit periodic amplification and attenuation, thus
appearing to be stable breathers. These oscillations are larger in one
component, although are present in both. Breathers in other DNLS systems
were reported in several works \cite{flach1997energy, johansson1998dynamics,
johansson2000growth, bambusi2009continuous}.

\section{Two-dimensional unstaggered-staggered lattices}

\label{sec3} In the higher-dimensional case, both theoretical \cite{vortex}
and experimental \cite{fleischer2003observation, fleischer2004observation}
work demonstrate that a variety of dynamics are possible for the DNLS. Here,
we extend the consideration of unstaggered-staggered modes to 2D lattices.
As in the 1D case, we first present the dynamical system and its Lagrangian
in subsection \ref{2dframe}. We then derive the VA in subsection \ref{2dva},
and present representative stable stationary two-component solitons in
subsection \ref{2dstab}.

\subsection{The 2D model and framework}

\label{2dframe} The natural 2D generalization of the system of coupled DNLS
equations (\ref{phipsi}) is
\begin{subequations}
\label{phipsi-2D}
\begin{equation}
i\frac{d}{dt}\phi _{m,n}=-\frac{1}{2}\left( \phi _{m+1,n}+\phi _{m-1,n}+\phi
_{m,n+1}+\phi _{m,n-1}-4\phi _{m,n}\right) -\left( \left\vert \phi
_{m,n}\right\vert ^{2}+\beta \left\vert \psi _{m,n}\right\vert ^{2}\right)
\phi _{m,n},
\end{equation}%
\begin{equation}
i\frac{d}{dt}\psi _{m,n}=-\frac{1}{2M}\left( \psi _{m+1,n}+\psi
_{m-1,n}+\psi _{m,n+1}+\psi _{m,n-1}-4\psi _{m,n}\right) -\left( \left\vert
\psi _{m,n}\right\vert ^{2}+\beta \left\vert \phi _{m,n}\right\vert
^{2}\right) \psi _{m,n}.
\end{equation}%
Solutions with unstaggered $\phi _{m,n}$ and staggered $\psi _{m,n}$
components and two chemical potentials, $\lambda $ and $\mu $, are sought
for as
\end{subequations}
\begin{equation}
\phi _{m,n}(t)=e^{-i\lambda t}u_{m,n},\quad \psi _{m,n}(t)=e^{-i\mu t}\left(
-1\right) ^{m+n}v_{m,n},  \label{stag-2D}
\end{equation}%
where real discrete fields $u_{m,n}$ and $v_{m,n}$ satisfy stationary
equations,
\begin{subequations}
\label{uv2D}
\begin{equation}\label{uv2Da}
\left( \lambda -2\right) u_{m,n}+\frac{1}{2}\left(
u_{m+1,n}+u_{m-1,n}+u_{m,n+1}+u_{m,n-1}\right) +\left( \left\vert
u_{m,n}\right\vert ^{2}+\beta \left\vert v_{m,n}\right\vert ^{2}\right)
u_{m,n}=0,
\end{equation}%
\begin{equation}\label{uv2Db}
\left( \mu -\frac{2}{M}\right) v_{m,n}-\frac{1}{2M}\left(
v_{m+1,n}+v_{m-1,n}+v_{m,n+1}+v_{m,n-1}\right) +\left( \left\vert
v_{m,n}\right\vert ^{2}+\beta \left\vert u_{m,n}\right\vert ^{2}\right)
v_{m,n}=0,
\end{equation}%
which can be derived from the corresponding 2D Lagrangian:
\end{subequations}
\begin{equation}
\begin{aligned} L & = \frac{1}{2}\sum_{m,n=-\infty }^{+\infty }\left[
-\frac{1}{2}\left( \left\vert u_{m+1,n}-u_{m,n}\right\vert ^{2}+\left\vert
u_{m,n+1}-u_{m,n}\right\vert ^{2}\right) +\frac{1}{2M}\left( \left\vert
v_{m+1,n}-v_{m,n}\right\vert ^{2}+\left\vert v_{m,n+1}-v_{m,n}\right\vert
^{2}\right) \right. \\ & \qquad\qquad\qquad\qquad \left. + \lambda
\left\vert u_{m,n}\right\vert ^{2}+\left( \mu -\frac{4}{M}\right) \left\vert
v_{m,n}\right\vert ^{2}+\frac{1}{2}\left\vert u_{m,n}\right\vert
^{4}+\frac{1}{2}\left\vert v_{m,n}\right\vert ^{4}+\beta \left\vert
u_{m,n}\right\vert ^{2}\left\vert v_{m,n}\right\vert ^{2}\right] .
\label{L2D} \end{aligned}
\end{equation}
In particular, note that \eqref{uv2Da} is the first variation of the Lagrangian \eqref{L2D} with respect to $u_{m,n}$, while \eqref{uv2Db} is the first variation of the Lagrangian \eqref{L2D} with respect to $v_{m,n}$.

\subsection{The variational approximation}

\label{2dva} In the 2D lattice framework, our first objective is to
construct two-component solitons with the fundamental onsite-centered
structure in both components. To apply the VA in this context, a practically
tractable ansatz may be taken as the 2D generalization of the exponential
one which was efficient in the application to the 1D DNLS equations \cite%
{MW1996,MW19962,VA14,we}:
\begin{equation}
u_{m,n}=Ae^{-P\left( |m|+|n|\right) },\quad v_{m,n}=Be^{-Q\left(
|m|+|n|\right) },  \label{2Dfund}
\end{equation}%
Considering ansatz \eqref{2Dfund} in the framework of linearized equations
at $n,m\rightarrow \pm \infty $, one finds $\lambda -2+2\cosh (P)=0$ and $%
\mu -\frac{2}{M}-\frac{2}{M}\cosh (Q)=0$. Solving for $P$ and $Q$ in terms
of $\lambda $, $\mu $, and $M$, we thus obtain
\begin{subequations}
\label{pq2D}
\begin{eqnarray}
P &=&\text{arccosh}\left( 1-\frac{\lambda }{2}\right) =\ln \left( 1-\frac{%
\lambda }{2}+\sqrt{-\frac{\lambda }{2}\left( 2-\frac{\lambda }{2}\right) }%
\right) ,~ \\
Q &=&\text{arccosh}\left( \frac{M\mu }{2}-1\right) =\ln \left( \frac{M\mu }{2%
}-1+\sqrt{\frac{M\mu }{2}\left( \frac{M\mu }{2}-2\right) }\right) \,.
\end{eqnarray}%
For $P$ and $Q$ to be real and positive, the allowed ranges of chemical
potentials $\mu $ and $\lambda $ are
\end{subequations}
\begin{equation}
\lambda <0\,,\quad \mu >\frac{4}{M}\,,  \label{><2D}
\end{equation}%
cf. its 1D counterpart (\ref{><}).

Real amplitudes $A$ and $B$ of ansatz (\ref{2Dfund}) are again treated as
variational parameters. In the 2D setting, a similar ansatz was applied to
the DNLS equation with the cubic-quintic onsite nonlinearity in Ref. \cite%
{CQ-2D}. VA was also used in a different context, for the rigorous proof of
the existence of discrete solitons as ground states in 1D \cite{variational}
and 2D \cite{MIW} settings. However, compound solitons built of unstaggered
and staggered components were not studied previously in any form.

It is also possible to construct 2D topological discrete solitons in which
at least one component is \textit{vortical} (hence, at least one of $u_{m,n}$
or $v_{m,n}$ is complex-valued), and this shall be the focus of the present
paper. For the single-component 2D DNLS equation, vortex-soliton solutions
were first constructed in Ref. \cite{vortex}. The initial ansatz for the
vorticity in one or both components of the onsite-centered soliton may be
taken as%
\begin{eqnarray}
u_{m,n} &=&Ae^{-P\left( |m|+|n|\right) },\quad v_{m,n}=B\left( m+in\right)
e^{-Q\left( |m|+|n|\right) },  \label{012d} \\
u_{m,n} &=&A\left( m+in\right) e^{-P\left( |m|+|n|\right) },\quad
v_{m,n}=Be^{-Q\left( |m|+|n|\right) },  \label{102d} \\
u_{m,n} &=&A\left( m+in\right) e^{-P\left( |m|+|n|\right) },\quad
v_{m,n}=B\left( m+i\sigma n\right) e^{-Q\left( |m|+|n|\right) },
\label{112d}
\end{eqnarray}%
where $P$ and $Q$ are as given by Eq. (\ref{pq2D}). \textit{Ans\"{a}tze} (%
\ref{012d}) and (\ref{102d}) represent compound solitons with the vorticity
embedded in one component, while in Eq. (\ref{112d}) both components are
assumed vortical, $\sigma =\pm 1$ being the \textit{relative vorticity} in
the two components. The amplitude shape of the stationary solutions is the
same for both $\sigma =+1$ and $-1$. In 2D continuum models, two-component
vortex solitons with opposite vorticities in the two components are called
\textit{counter-rotating} two-component vortices \cite{counter}, alias
states with \textit{hidden vorticity} \cite{Nal}, if the total angular
momentum is zero. In that case, a nontrivial issue is stability of such
compounds, which may be very different from that of their co-rotating
counterparts.

Existing studies of modulational instability in 2D DNLS lattices also
suggest that plane and solitary waves are often modulationally unstable to
small-wavenumber perturbations \cite{kevrekidis2001comparison,
hudock2003discrete}. The development of instability may lead to formation of
highly localized structures, such as solitons or other localized states.
Akin to the 1D case, the 2D solitons of the form given by Eqs. \eqref{2Dfund}%
, \eqref{012d}-\eqref{112d} are expected to have narrow envelopes, with
larger values of $P$ and $Q$, while broad envelopes corresponding to smaller
values of $P$ and $Q$ are likely to be unstable.

While in 1D we showed that only fundamental-fundamental, fundamental-twisted, and
fundamental-fundamental soliton pairs exist in unstaggered-staggered
lattices, in the 2D case we have found no asymmetric vortical-fundamental
pairs, so we only present the VA for fundamental-fundamental and
vortical-vortical pairs.

\subsubsection{Fundamental unstaggered - fundamental staggered pairs}

Here we construct 2D fundamental solitons of the unstaggered-staggered type,
based on ansatz \eqref{2Dfund}. Substituting it in Lagrangian (\ref{L2D})
and carrying out the summation, we arrive at the effective Lagrangian:
\begin{eqnarray}
2L_{\mathrm{eff}} &=&-2A^{2}\left( \coth P\right) \tanh (P/2)+\frac{2}{M}%
B^{2}(\coth Q)\tanh (Q/2)+\lambda A^{2}\coth ^{2}P+\left( \mu -\frac{4}{M}%
\right) B^{2}\coth ^{2}Q  \notag \\
&&+\frac{A^{4}}{2}\coth ^{2}(2Pp)+\frac{B^{4}}{2}\coth ^{2}(2Q)+\beta
A^{2}B^{2}\coth ^{2}(P+Q)\,,  \label{Leff2d}
\end{eqnarray}%
which gives rise to the Euler-Lagrange equations,
\begin{equation}
A^{2}\coth ^{2}(2P)A^{2}+\beta B^{2}\coth ^{2}(P+Q)=2(\coth P)\tanh
(P/2)-\lambda \coth ^{2}P\,,  \label{vara2d}
\end{equation}%
\begin{equation}
\beta A^{2}\coth ^{2}(P+Q)+B^{2}\coth ^{2}(2Q)=-\frac{2}{M}(\coth Q)\tanh
(Q/2)-\left( \mu -\frac{4}{M}\right) \coth ^{2}Q\,.  \label{varb2d}
\end{equation}
As $P$ and $Q$ are determined already from \eqref{pq2D}, the system \eqref{vara2d}-\eqref{varb2d} determines the unknown amplitude parameters $A$ and $B$. It is clear that a physically relevant solution pair, with $A^{2}>0$, $%
B^{2}>0$ does not exist for $\beta >0$, which is a natural consequence of
the fact that we are looking for the unstaggered-staggered complex. However,
physical solutions may exist for $\beta <0$, i.e., with opposite signs of
the SPM and XPM onsite terms in Eqs. (\ref{phipsi}). Further, the system of
Eqs. \eqref{vara2d} and \eqref{varb2d}, considered as a system of linear
equations for $A^{2}$ and $B^{2}$, becomes degenerate if
\begin{equation}
\coth ^{2}(2P)\coth ^{2}(2Q)=\beta ^{2}\coth ^{4}(P+Q)\,.  \label{degen2d}
\end{equation}

\subsubsection{Vortical unstaggered - vortical staggered pairs}

The most sophisticated compound mode is built of vortical modes in both the
unstaggered ($u$) and staggered ($v$) components, as per ansatz (\ref{112d}%
). Substituting the ansatz in Lagrangian (\ref{L2D}), we obtain the
respective effective Lagrangian,
\begin{eqnarray}
2L_{\mathrm{eff}} &=&-A^{2}\eta (P)+\frac{B^{2}}{M}\eta (Q)+\lambda
A^{2}\coth ^{2}(P)(\sinh (P))^{-2}+\left( \mu -\frac{4}{M}\right) B^{2}\coth
^{2}(Q)(\sinh (Q))^{-2}  \notag \\
&&+\frac{A^{4}}{2}\zeta (2P)+\frac{B^{4}}{2}\zeta (2Q)+\beta A^{2}B^{2}\zeta
(P+Q)\,,  \label{Leff112d}
\end{eqnarray}%
where we have defined
\begin{equation}
\eta (\alpha )\equiv \frac{(\sinh (\alpha ))^{-1}\coth (\alpha )}{\cosh
(\alpha )+1}+\frac{1}{2}\tanh (\alpha /2)\coth (\alpha )(\sinh (\alpha
))^{-2}\,,  \label{eta2d}
\end{equation}%
\begin{equation}
\zeta (\alpha )\equiv \frac{1}{2}\coth ^{2}(\alpha )(\sinh (\alpha
))^{-4}\left( \cosh (2\alpha )+6\right) \,.  \label{chi2d}
\end{equation}%
From here we derive the corresponding Euler-Lagrange equations,
\begin{equation}
A^{2}\zeta (2P)+\beta B^{2}\zeta (P+Q)B^{2}=\eta (P)-\lambda \coth
^{2}(Q)(\sinh (P))^{-2}\,,  \label{var11a2d}
\end{equation}%
\begin{equation}
\beta A^{2}\zeta (P+Q)+B^{2}\zeta (2Q)B^{2}=-\frac{1}{M}\eta (Q)-\left( \mu -%
\frac{4}{M}\right) \coth ^{2}(Q)(\sinh (Q))^{-2}\,.  \label{var11b2d}
\end{equation}
Again, as $P$ and $Q$ are determined already from \eqref{pq2D}, the system \eqref{var11a2d}-\eqref{var11b2d} determines the unknown amplitude parameters $A$ and $B$. Since $\zeta (\alpha )>0$ for $\alpha >0$, in the present case we conclude,
as in the 1D case, that a solution pair with $A^{2}>0$, $B^{2}>0$ does not
exist for $\beta >0$, but relevant solutions may exist at $\beta <0$. The
system of Eqs. \eqref{var11a2d} and \eqref{var11b2d} becomes degenerate if
\begin{equation}
\zeta (2P)\zeta (2Q)=\beta ^{2}\zeta (P+Q)^{2}\,.
\end{equation}

\subsection{Stable 2D solution pairs}

\label{2dstab} As in the case of the 1D lattice, in the 2D case we use the
VA \textit{ans\"{a}tze} as initial guesses for solving Eqs. \eqref{uv2D},
and then simulate the ensuing evolution in the framework of Eq. %
\eqref{phipsi-2D}. The functions and tolerances used are the same as in the
1D setting, although we here restrict the lattice to smaller sizes, for
computational reasons. As the lattice is smaller, we simulated the evolution
of the solutions for shorter times than in 1D, to avoid artifacts caused by
radiation reflecting from the domain's boundaries. Nevertheless, concluding
if the localized solution remain stable in the respective time interval (and
for longer times on increasingly large lattices), it is possible to
conjecture that such solutions will remain stable indefinitely.

\begin{figure}
\centering
\includegraphics[width=0.4\textwidth]{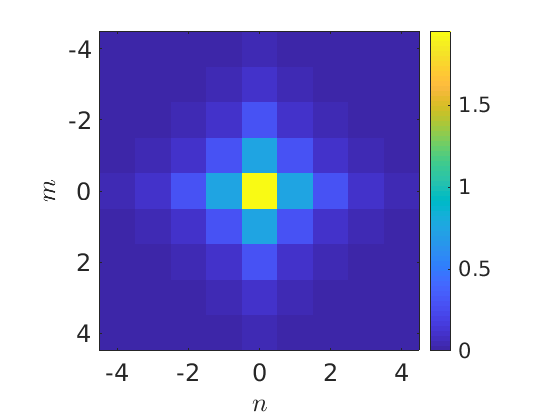} \includegraphics[width=0.4%
\textwidth]{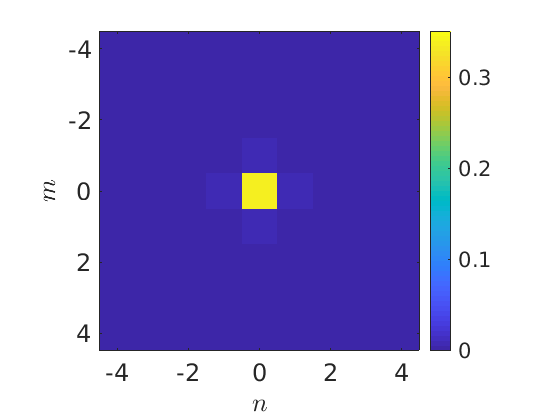}
\par
(a) \hspace{6cm} (b)
\par
\includegraphics[width=0.4\textwidth]{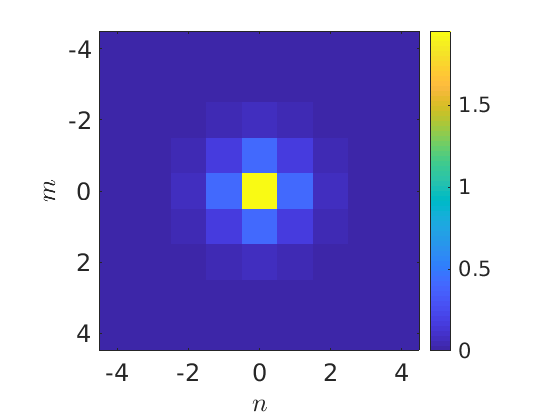} \includegraphics[width=0.4%
\textwidth]{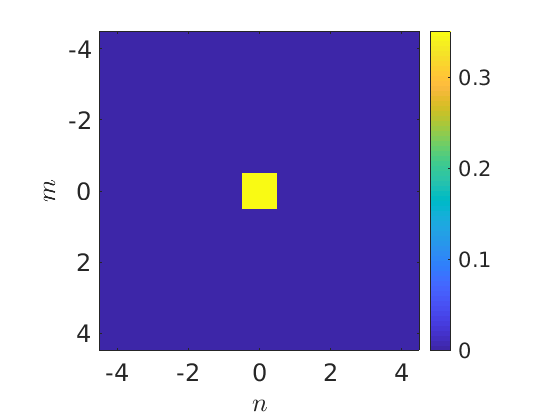}
\par
(c) \hspace{6cm} (d)
\caption{2D two-component solitons of the fundamental-fundamental type for
parameters $\protect\beta =-10$, $\protect\mu =40$, $\protect\lambda =-1$, $%
M=1$. Discrete fields $|u_{m,n}|$ and $|v_{m,n}|$ are plotted in (a,c) and
in (b,d), respectively, with (a,b) and (c,d) representing, respectively, VA
ansatz \eqref{2Dfund} and the numerical steady- state solutions of Eq.
\eqref{uv2D}. }
\label{2dfundamental}
\end{figure}

In Figure \ref{2dfundamental} we display an example of a
fundamental-fundamental soliton pair. In the course of the evolution,
staggered component $\psi $ emits some radiation, and then remains localized
with a steady absolute-value profiles. In contrast, the unstaggered
component $\phi $ emits no radiation at all. We conclude that the VA for
these solutions is qualitatively accurate, although the numerically observed
soliton is more localized in the $u_{m,n}$ component than the VA predicts.
For these solutions, we used a lattice of size $61$ by $61$ and could
observe a stable soliton up to $T=100$. Thus, complementing what was shown
in Ref. \cite{we} for the 1D case, we find that there exist stable
fundamental-fundamental soliton pairs in the 2D unstaggered-staggered
lattice.

\begin{figure}
\centering
\includegraphics[width=0.4\textwidth]{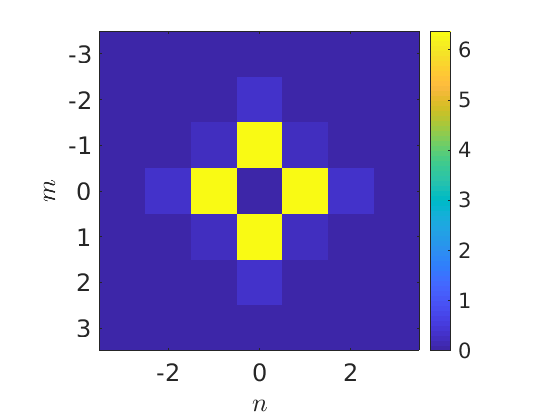} \includegraphics[width=0.4%
\textwidth]{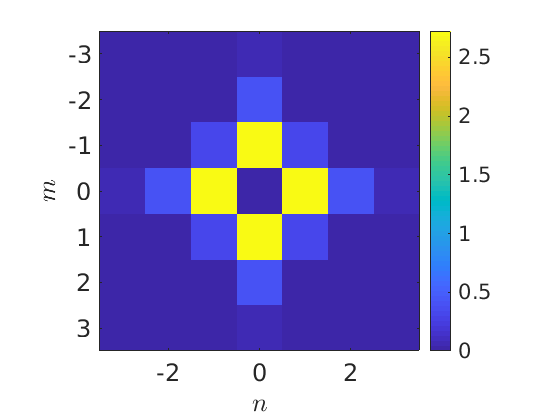}
\par
(a) \hspace{6cm} (b)
\par
\includegraphics[width=0.4\textwidth]{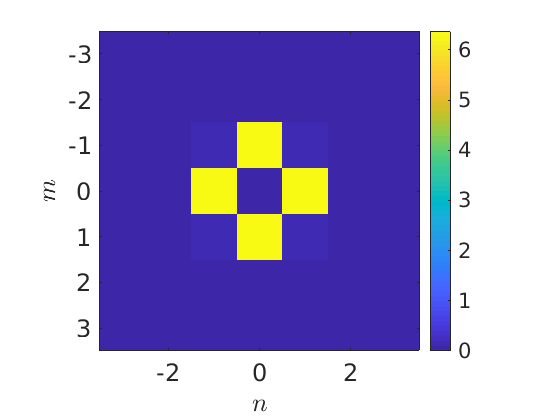} \includegraphics[width=0.4%
\textwidth]{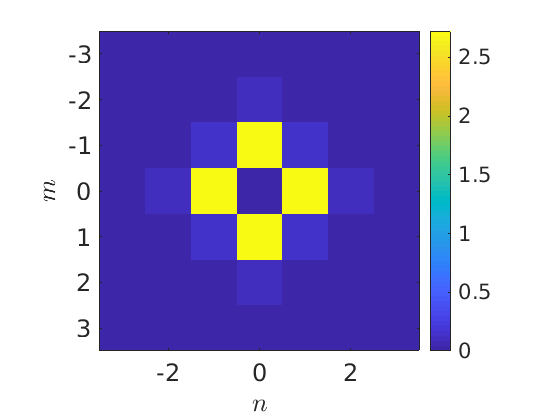}
\par
(c) \hspace{6cm} (d)
\caption{2D vortical-vortical solution pairs with parameters $\protect\beta %
= -0.5$, $\protect\mu = 15$, $\protect\lambda = -35$, $M=1$, with $\protect%
\sigma=1$ used in the VA ansatz. We plot $|u_{m,n}|$ in (a,c) and $|v_{m,n}|$
in (b,d), with (a,b) corresponding to the VA ansatz \eqref{112d} and (c,d)
the numerical steady state solutions of \eqref{uv2D}. }
\label{2dvortuvpos}
\end{figure}

We were unable to find any parameter values which admitted solitons of the
mixed staggered forms corresponding to Eqs. \eqref{012d} or \eqref{102d}.
Parameters permitting the existence of such solutions should be quite
sparse, if they exist at all. This is consistent with the above-mentioned
difficulty in finding\ 1D stable fundamental-twisted solitons, and the
non-existence of twisted-fundamental pairs. In contrast, we have found pairs
of vortical solitons, many of which are stable. We produce an example of
such soliton pairs in Fig. \ref{2dvortuvpos}, which are stable for both co-
and counter-rotating vortical pairs (i.e., with~$\sigma =\pm 1$). Again,
some radiation is emitted, in the course of the short-time relaxation, only
by the staggered component, and the resulting localized solution persists
after the completion of the relaxation. In this case too we conclude that
the VA is a reasonable fit to the form of the numerically computed
solutions. Here we restricted the lattice to a size of $41$ by $41$, and
ensured the stability until $T=50$. Nonetheless, we expect the solitons
shown in Fig. \ref{2dvortuvpos} to persist indefinitely. The finding of the
stable vortical-vortical soliton pairs on the unstaggered-staggered lattice
adds to previously known results for discrete vortex solitons on the usual
2D DNLS lattice \cite{vortex}.

\section{Conclusions}

\label{sec4} Extending the analysis of the recently introduced system of
nonlinearly coupled DNLS equations with unstaggered and staggered components
(which requires opposite signs of the SPM and XPM nonlinearities---a
situation possible in binary BEC), we have elaborated families of 1D
discrete solitons with a single twisted or both twisted components,
complementing the earlier work on fundamental soliton pairs on
unstaggered-staggered lattices \cite{we}. Analytical solutions for the
discrete solitons are constructed by means of the VA\ (variational
approximation). Similar to the recently studied family of fundamental
solitons in this system \cite{we}, we have found that the twisted solitons
produced by the VA are often stable when they are narrow, and unstable (or
nonexistent in simulations) if at least one component is wide. We find that
the VA is in the best agreement with numerical simulations when the solution
pairs are symmetric, such as the fundamental-fundamental and twisted-twisted
ones. As for asymmetric twisted-unstaggered--fundamental-staggered pairs,
while stable solutions can be found numerically, they do not well agree with
the VA, highlighting a limitation of the VA in that case. On the other hand,
asymmetric fundamental-unstaggered--twisted-staggered pairs are not
predicted by the VA, and were not found in simulations either, therefore we
conjecture that they do not exist. Through numerical simulations, we have
determined the long-time evolution of initial steady states, with stable
solutions maintaining their shape (sometimes after giving off a small amount
of radiation, as they adjust to a true stable soliton), and unstable
solutions decaying or exhibiting transient dynamics. In addition to the
decay of both wave functions due to the instability, other unstable soliton
initial conditions were observed to evolve into breathers or lead to the
amplitude death (vanishing) of one wave function (with the other component
persisting as a soliton).

Additionally, we have considered the extension of the unstaggered-staggered
formulation to 2D lattices, which was not considered previously, and
constructed both fundamental solitons and vortical ones, producing
representative stable solutions for each case. Stable 2D two-component
solitons of the asymmetric vortical-fundamental or fundamental-vortical have
not been found, in agreement with the fact that, in the 1D case, twisted
unstaggered-fundamental staggered pairs exist in a limited area of the
parameter space, while the pairs of the symmetric types
(fundamental-fundamental and twisted-twisted ones) are more common. Thus,
the twisted-twisted (1D) and vortical-vortical (2D) pairs are found to be
more robust than their asymmetric counterparts.

The solutions that we have found to be stable often correspond to narrow
envelopes, consistent with results known form previous works. In particular,
this finding is in agreement with the modulational-instability analysis for
plane waves in related coupled NLS and complex Ginzburg-Landau systems in
both continuum and discrete settings \cite{wabnitz1988modulational,
kobyakov1998dark, abdullaev2002modulational, rapti2004parametric,
baizakov2009modulational, alcaraz2010modulational}, including two and three
spatial dimensions \cite{hudock2003discrete,van2018coupled}. Indeed,
perturbations with small wavenumbers often lead to modulational instability
of plane waves \cite{newton1987stability,tan2001stability}, which frequently
results in the creation of highly localized structures, including solitons
and other localized states \cite{kivshar1992modulational}. Furthermore, the
direct modulational-instability analysis, applied to solitons in other
related lattice systems, likewise suggests that modes with narrow envelopes
tend to be stable \cite{hudock2003discrete}, while wider ones fail to
persist in the course of time evolution.

The 1D and 2D solutions\ that we have obtained here are novel in the context
of unstaggered-staggered lattices, and they may help to motivate future
theoretical and experimental work in BEC and optics. Regarding theoretical
extensions, there are a number of ways these results may be extended. We
remark that \textit{intersite-centered} 2D solitons and vortices may be
considered too, although they are expected to be much less stable \cite%
{Panos}. The ansatz for intersite-centered modes can be obtained from Eqs. %
\eqref{2Dfund} and \eqref{012d}-\eqref{112d} by replacing $\left\{
m,n\right\} \rightarrow \left\{ m-1/2,n-1/2\right\} $. It may also be
interesting to consider solutions on periodic domains, such as a ring or
torus. Work in this direction was reported in Refs. \cite%
{ring,ring2,ring3,ring4}. As the soliton tails decay fairly rapidly, there
may be little difference in the form of the solutions if the ring or torus
is large enough, while, as they are made smaller, one may expect curvature
effects to come into play. Finally, a more systematic treatment of some
unsteady structures found here, such as breathers and the dynamics leading
to \textquotedblleft amplitude death", may be explored in more depth.

\end{document}